\documentclass{article}
\usepackage[utf8]{inputenc}
\usepackage{hyperref} 
\usepackage{graphicx}
\usepackage{multicol,float}
\usepackage{algorithm}
\usepackage{algorithmic}

\title{Approximate Strategyproofness in Large, Two-Sided Matching Markets}
\author{Lars L. Ankile\footnote{Corresponding author: lars.ankile@gmail.com}, Kjartan Krange, Yuto Yagi \\ {\em\small Harvard College}\\{\em\small Cambridge, MA}}

\date{}

\begin{document}

\maketitle

\begin{abstract}
An approximation of strategyproofness in large, two-sided matching markets is highly evident. Through simulations, one can observe that the percentage of agents with useful deviations decreases as the market size grows. Furthermore, there seems to be a strong connection between the length of preference order lists, the correlation of agent preferences, and the approximation of strategyproofness. Interestingly, approximate strategyproofness is reached easier with a shorter length of preference orders and higher preference correlation. These findings justify the use of the deferred acceptance algorithm in large two-sided matching markets despite it not being strategy-proof.

\end{abstract}

\section{Introduction}
\begin{multicols}{2}
Approximate strategyproofness is observed in large matching markets where the percentage of agents who can usefully deviate is small. This phenomenon is further explored in \cite{roth}. It presents how the labor-market matching between newly graduated physicians and hospitals benefit from the discovery that \textit{"Opportunities for strategic manipulations, are surprisingly small"} i.e. they show how approximate strategyproofness makes their matching algorithm more robust in a large market. Clearly, the implications of the property could, in many cases, be important when deciding how to implement large matching markets, for example, by allowing market designers to trust the agents to report truthfully even though the DA mechanism is not technically strategy-proof.

In this paper, we run multiple simulations to analyze how and when one can approximate strategyproofness. Furthermore, we aim to simulate different amounts of correlation between the agents' preferences in the market. In researching these properties, we hope to get new insights into when approximate strategyproofness can reliably mitigate gaming of the system and for what market size, length of preference lists, and preference correlation it appears effective to a lesser extent.  

We hypothesize that we will witness a near strategyproofness for sufficiently large markets and that the length of the reported preference ordering and the amount of correlation will affect the approximation to strategyproofness. All source code can be found on the \href{https://github.com/weird-foreign-guys/large_matching_market}{\textbf{GitHub-repo}}.
\end{multicols}

\section{Theory}
\begin{multicols}{2}
 Our study is based on several pieces of existing research. The following is a list of theorems with a description of how they apply to our paper:

\begin{itemize}
    
     \item \textit{Simplicity, Dynamic Stability, and Robustness are some of many wanted properties of strategyproofness} \cite{lubin}
     \textbf{Thus, we want to have a strategy-proof mechanism design}
    \item \textit{Theorem 12.5 Truthful reporting is a dominant strategy for students in the student-proposing DA mechanism} \cite{parkes}
    
    \item \textit{Theorem 12.7 No mechanism for two-sided matching is both stable and strategy-proof} \cite{parkes}
    
    Knowing that deferred acceptance returns a stable matching, and that the proposing side has truthful reporting as its dominant strategy, we can deduce that in general there exists useful deviations, and that these will only exits for the agents that are being proposed to.
    
    \textbf{Thus, we only have to look for deviations among agents being proposed to}
    
    \item \textit{In simple markets, [...] all successful manipulations can [also] be accomplished by truncations} \cite{roth}
    
    This result is highly useful in the implementation of our simulation. Instead of looking for $k!$ possible permutations of preferences, for preference orders of length $k$, it is sufficient to truncate the preference order once at every position for each participant on the receiving side. This makes an intractable problem tractable. \textbf{Thus, we only need to check for truncation deviations}.
    
    The theorems mentioned strategyproofness desirable. Furthermore, in the search for approximated strategyproofness, we can conclude that searching for deviations only in the form of truncation on the receiving side of the market can be done without loss of generality. This reduces the computational complexity immensely.

\end{itemize}
\end{multicols}

\newpage
\section{Method}
\begin{figure}
    \centering
    \includegraphics[width=1.0\textwidth]{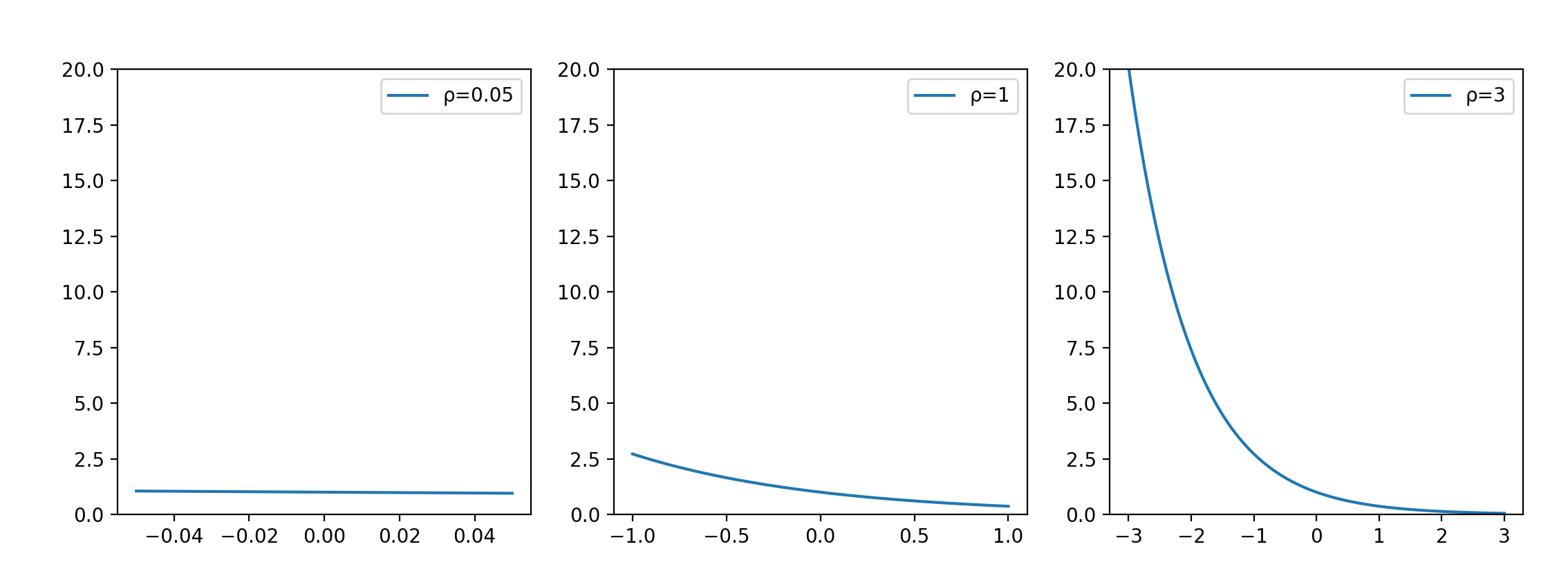}
    \caption{Plots of the three probability distributions we used for modeling different levels of correlation between agents' preferences. Left is the uncorrelated distribution, middle is the moderate, and right is the high correlation.}
    \label{fig:rho}
\end{figure}
\begin{multicols}{2}
In short, our simulation generates matching markets for a given amount of preference correlation and length of preference orders, gradually increasing the market size, and for each market size, counts the number of agents who can usefully deviate.

Our algorithm can be divided into three parts: \textit{(i) Preference Generation}, \textit{(ii) Deferred Acceptance}, \textit{(iii) Counting Deviations}.

\subsection{Preference Generation}


For our results to be valid we need to realistically model agent preferences in a way that would resemble the distribution over preferences in a real market. An important aspect we need to capture is that some market participants tend to be popular among all agents on the other side, e.g. some men are usually attractive in most women's eyes, which makes them more likely to appear high on women's hypothetical preference orderings. We found that a distribution defined by the inverse exponential function would give us a fitting representation. We model three specific market scenarios with differing level of correlation between agents' preferences: (1) almost no correlation, (2) moderate correlation, and (3) high correlation. In figure \ref{fig:rho}, one can see a plot of the three different distributions used, respectively.

Based on these distributions we first generated preference orderings for one side of the market, where they were given a $k$ length preference list drawn at random according to the distributions described above. Since it e.g. does not makes sense for a hospital to have preferences over students they have not interviewed, we chose to base the proposal-receiving side's preferences on the preferences of the proposers. This will also make sure we have as high match ratio as possible. We allowed the proposal receiving side to have a variable length of preferences. E.g. if every male was allowed to have preference orderings of length 10, and there are 5 males who have one specific female among their preferences, that female will have a preference ordering of length 5. With everyone's preferences in order we are ready to proceed to the actual matching.

\subsection{Deferred Acceptance}
In order to find stable matches, we implemented the deferred acceptance algorithm which is known to produce stable matches. This algorithm takes incomplete preference orders and interprets it as the agent preferring to be unmatched to being matched agents not on their list. E.g. suppose $m$'s preference over $f_1, f_2, f_3$ given as $f_1 \succ_{m} f_3 $. Then this will be interpreted as $f_1 \succ_{m} f_3 \succ_{m} \emptyset \succ_{m} f_2$. The following is the pseudo code of our deferred acceptance algorithm. We had to implement this algorithm from the bottom up, instead of using libraries, because most existing code did not handle truncations and incomplete preference orders.

\end{multicols}

\begin{algorithm}
\caption{\textsc{DefferedAcceptance}}\label{euclid}
\begin{algorithmic}
\STATE Initialize p in proposers and r in recipients unmatched

\WHILE{exists p that is unmatched and has someone yet to propose to}
    \FOR {p in unmatched proposers}
    \STATE r = most preferred r which p has not proposed yet
        \IF{p is not in r's preference list}
            \STATE $break$
        \ELSIF {r is unmatched}
            \STATE (p, r) become matched
        \ELSE 
            \IF{r prefers p to current match}
            \STATE{(p, r) become matched \\
            r's previous match p' becomes unmatched}
            \ELSE
            \STATE {no change}
        \ENDIF
        \ENDIF
    \ENDFOR
\ENDWHILE
\end{algorithmic}

\end{algorithm}

\newpage
\begin{multicols}{2}
\subsection{Counting Deviations}

This paper is trying to find a connection between $D(n)$, i.e. the number of agents with useful deviations, and $n=$, the number of agents on one side of the market. Subsection 3.3 describes the calculation of $D(n)$.

As presented in section \textit{2 Theory}, for the stable matchings produced by a given mechanism, misreports in the form of truncation deviation by the receiving side of the market is the only possibility for useful deviations. For simplicity, we call a recipient \textit{female} in this subsection (without intention of being hetero-normative). When our simulation returns a stable matching based truthful reports (given a preference generation dependent on $\rho$, $k$, and $n$) we analyze this matching in the following way:

\begin{enumerate}
    \item \textnormal{While there are females not analyzed, select a female.}
    \item \textnormal{Given a female, truncate her last preference order entry.}
    \item \textnormal{Based on this new preference order, run deferred acceptance and check if this female benefits.}
    \item \textnormal{Repeat step 2 and 3 until deviation is found (if so set $D(n) = D(n)+1$) or preference list is of length one (in this case mark female as analysed and go to step 1). }
\end{enumerate}

We are counting how many females have at least one useful misreport. The $D(n)$ returned will be logged with the corresponding $n=\textnormal{market size}$, $k=\textnormal{preference length}$ and $\rho=\textnormal{preference correlation}$. Giving us our final results.
\end{multicols}

\section{Results}
In this section we briefly present the results from our extensive simulations in the form of a series of plots. This required 300+ CPU-hours.

\begin{figure}[H]
    \centering
    \makebox[\textwidth][c]{\includegraphics[width=1.4\textwidth]{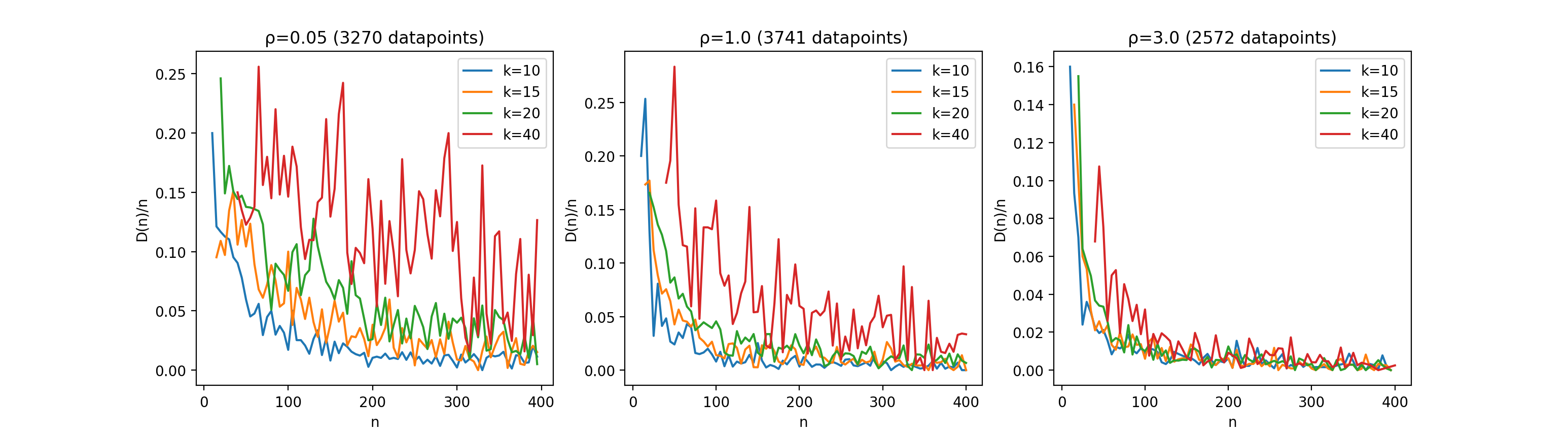}}%
    \caption{Simulations for $\rho = 0.05, 1.0, 3.0$ (from left to right) each with $k = 10, 15, 20, 40$.}
    \label{fig:overview}
\end{figure}

\begin{figure}[H]
    \centering
    \makebox[\textwidth][c]{\includegraphics[width=1.3\textwidth]{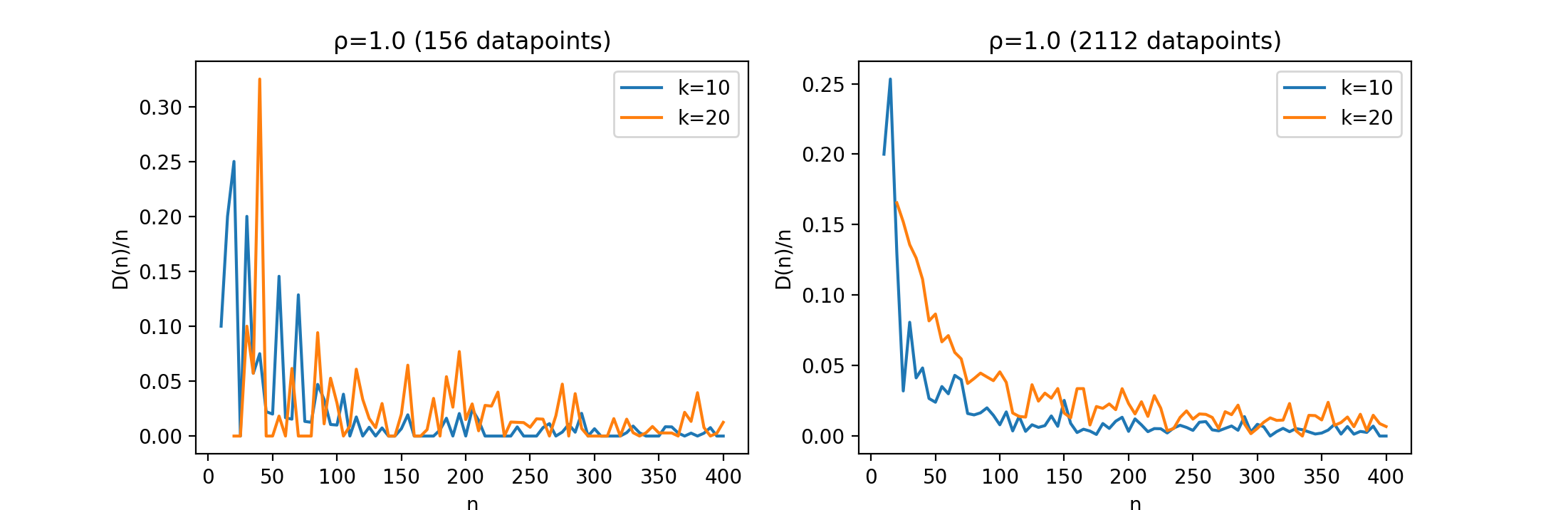}}%
    \caption{Plot for 156 data points to the left and 2112 data points to the right. With increasing data density, curves converge towards clearer separation.}
    \label{fig:data-density}
\end{figure}

\begin{figure}[H]
    \centering
    \includegraphics[width=0.85\textwidth]{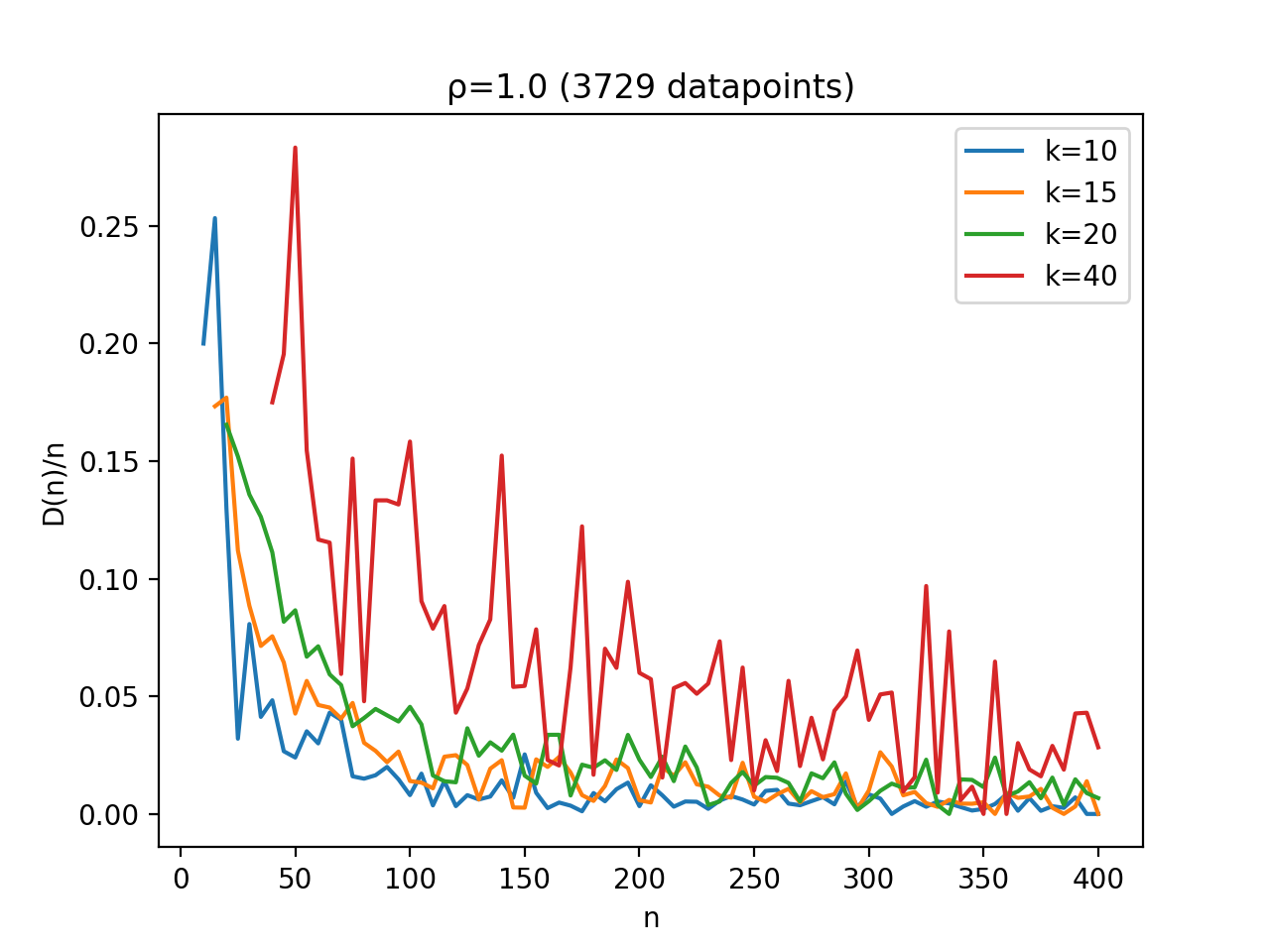}
    \caption{Plot for $\rho=1.0$ and for multiple values of $k$. Observe that longer preference orderings tend to make the opportunities for strategic behavior larger.}
    \label{fig:vary-k}
\end{figure}

\begin{figure}[H]
    \centering
    \includegraphics[width=0.85\textwidth]{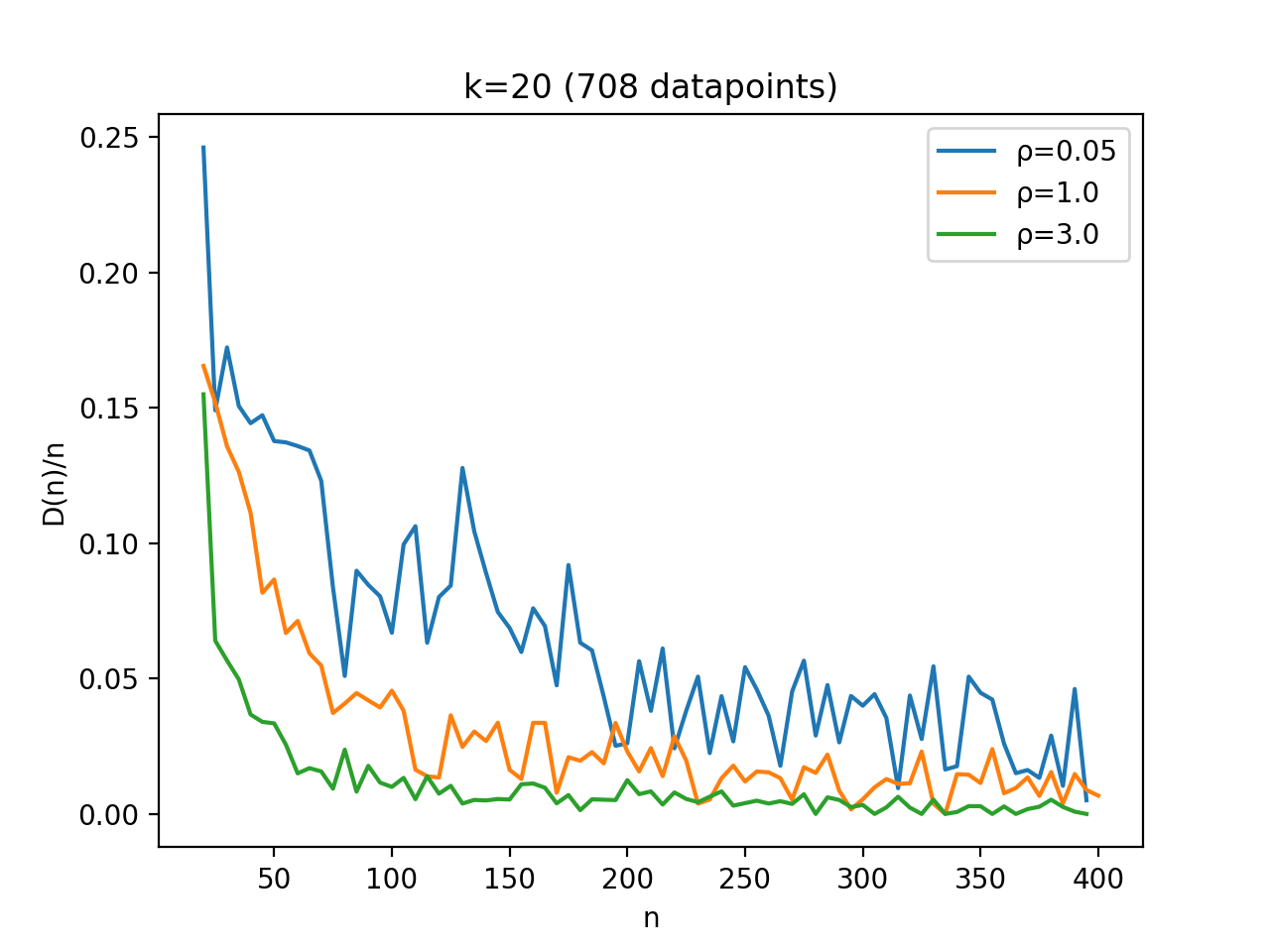}
    \caption{Plot for $k=20$ for a range of different $\rho$. Observe that correlation between agents' preferences tend to increase the amount of opportunities for strategic behavior.}
    \label{fig:vary-r}
\end{figure}

\begin{figure}[H]
    \centering
    \makebox[\textwidth][c]{\includegraphics[width=1.3\textwidth]{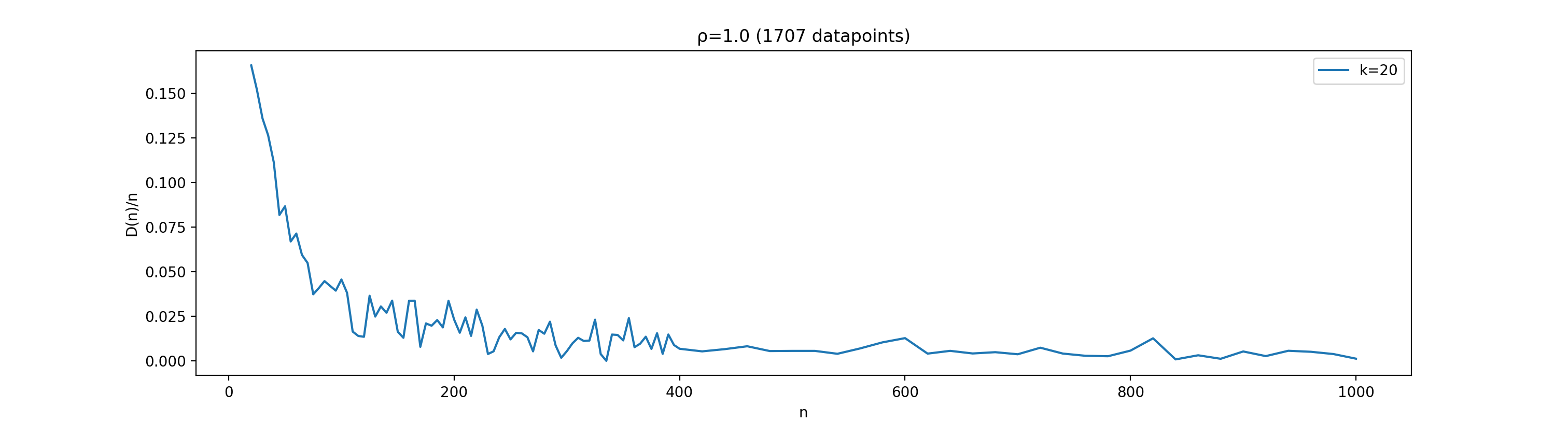}}
    \caption{Plot for $k=20$, $\rho=1$, for markets of size up to $n=1000$. We assume that other values for $k$ and $\rho$ will produce proportional results in the limit.}
    \label{fig:n1000}
\end{figure}

\newpage
\section{Discussion}

\begin{multicols}{2}
\subsection{General Findings}

There are two main types of simulations performed. The first is for four different preference lengths, $k=10, 15, 20, 40$, and three amounts of preference correlations, $\rho = 0.05, 1.0, 3.0$, with a market size $n$, running from 5 to 400 (figure \ref{fig:overview}). This illustrates the effect of different $k$ and $\rho$ on the approximate strategyproofness. The second type shows how, as the market grows even larger (up to $n=1000$), $D(n)/n$ converges to 0 (figure \ref{fig:n1000}). We show this while keeping $k$ and $\rho$ constant at 20 and 1.0, respectively. Because we lack the resources to run the same simulation for the other values of $k$ and $\rho$, we chose $k=20$ and $\rho=1.0$ as representative values. We assume that the other curves would also tend towards 0.

As $k$ and $n$ grows, the amount of computation required grows relatively quickly. Our deferred acceptance implementation has a running time of $\mathcal{O}(n^2)$. We run the DA for $n$ recipients with an average preference length $k$, leaving our total simulation at an average running time of $\mathcal{O}(n^3k)$. An unrealistic, but definite worst case would be $\mathcal{O}(n^4)$, for $k=n$.

As a general finding, the results show that the market quickly approaches strategyproofness when the markets grow from a size of $n=10$ up to around $n=100$. In fact, in this segment of $n$, given a high $\rho$ the percentage of agents with useful deviations goes from around 16\% to a mere 0.01\%. As the market grows even larger, the percentage of agents with useful deviations approaches zero, but the speed of convergence slows down (see e.g. figure \ref{fig:vary-r}).

Our last figure shows how with an increasing number of simulation runs, the data plot gradually becomes smoother with less random spikes in $D(n)/n$. Since the generation of the preference orders are partially based on randomness, having the data based on multiple simulations give us model results gradually approaching an average value. The random based algorithm design, makes our results more easily applicable to the real world as they are based on a huge set of different kind of matching market situations. The plot gradually converges towards a monotone decreasing function. This leads us to conclude that approximate strategy proofness generally exists in large markets. We cannot, however, always rule out the existence of the odd possibility for deviation for some agent.

\end{multicols}

\begin{multicols}{2}
\subsection{Effect of Varying Preference-Order Length}

In figure \ref{fig:overview} we see how the three plots: $k=10$ in blue, $k=15$ in orange, and $k=20$ in green. The important thing to note is that in general the plots of $D(n)/n$ for a given $n$ increases with $k$. Intuitively this makes sense as a larger $k$ quite simply gives more preferences to truncate away, i.e. more possible deviations. For a market designer, the general advice would be that a smaller preference reporting allowed restricts the amount of beneficial misreports available for all agents in total. 

\end{multicols}

\begin{multicols}{2}
\subsection{Effect of Varying the Preference Correlation}

 We note that an increasing correlation, $\rho$ ($\rho$=0.05, $\rho$=1 and $\rho$=3), leads to smaller ratio of agents who can usefully deviate ($D(n)/n$). This allows market designers with a knowledge of low correlation to expand the market to secure approximate strategy proofness. As for the reason why stronger correlation leads to less possibility for useful deviation, having more agents interested in the same matches makes the rejection chain shorter, thus making it unlikely to return to the deviator.

\end{multicols}
\section{Applications and Improvements}

\begin{multicols}{2}
As discussed, a matching market designer has to choose between a stable and a strategy proof matching. Parkes and Seuken (2019\cite{parkes}) argue that unstable markets have a tendency to unravel (possibly with catastrophic consequences) which makes stability an arguably more important property than strategyproofness. However, as shown in this paper, large market can gain approximate strategyproofness while maintaining the stability of the outcome.

In DA, the proposal-receiving side gets their least preferred, achievable match. In some cases it might actually be more important for organizing party to get the better matching. An example could be a market trying to match medical students to residencies at hospitals. It is potentially better if the hospital-optimal matching was achieved instead of the student-optimal one. This is because the needs for students with specific talent and knowledge at the hospitals might be more important for the society at large the individual wants of the students.

The results in this paper are likely applicable to the real world. We have seen that in two-sided matching markets it quickly becomes difficult to find useful deviations for the receiving side. This effect is particularly evident when we have relatively short preference orderings and a high correlation between the preferences of the agents in the market. Both of these assumptions we think are reasonable in real-world applications.

As a familiar instance, Norway has a coordinated admissions system for almost every institution of higher education, \textit{Samordna opptak}. There, students are ranked according to their high school grades, and students are allowed to create a strict preference order over at most 10 schools. Here we see that every agent on the proposing side has a preference ordering of length at most 10, while every recipient has a preference ordering over all students who have them on their preference ordering. Furthermore, we know from open statistics that a select few schools are extremely popular, while many schools are not as popular and struggle to fill up their classes. This effect is probably common, and is well captured by our medium or high correlation distribution. These observations, together with the fact that the market is very large (tens of thousand of agents), make us believe that this market would potentially reach strong approximate strategyproofness. 

As a last point, we want to argue that even though our simulation is implemented with an even-sized pool of proposers and proposed agents, a higher $\rho$ models a matching market with a different sized sides. When $\rho$ is large almost all proposed agents are interested in a select few proposers. This is close to a real world situation where there are few proposers compared to the proposed side. As such, an increasing $\rho$ could be used to model a situation with an increasingly large amount of agents on the proposed side of the market relative to the proposers.

\end{multicols}

\section{Potential Issues and Next Steps}
\begin{multicols}{2}

The following are some points of issues and possible future implementations: 

\begin{itemize}
    \item Our simulation assumes an equal number of proposer and recipients, and have not modeled the case where there are more proposers than recipients.
    \item If we were to use our code to model more varied markets, we would gain a lot from having a more highly optimized algorithm. 
    \item We have modelled one-to-one two sided matching market, but there are many other form of matching that might require analysis. They include matching with couples, one-to-many matchings, and many-to-many matchings. Next step would be to expand our model to include these different matching markets.
    \item This paper has given a empirical proof on the existence of approximate strategy proofness in large two sided matching markets. The next step would be to provide a theoretical proof as to how and why this property arise in large markets.
\end{itemize}
\end{multicols}

\section{Conclusion}
We have found approximate strategyproofness to be prevalent in large markets. Our models show that the percentage of agents with useful deviations decreases most rapidly between a market size of 10 and 100. Approximate strategyproofness decreases with the length of preference orders and increases with preference correlation. Lastly, there are a number of real world applications, like the Norwegian university-application process, where these findings are of interest. There still exists open issues that we could address in future research.

\section*{Statements \& Declarations}

\textbf{Funding:} The authors declare that no funds, grants, or other support were received during the preparation of this manuscript.

\textbf{Competing interests:} The authors have no relevant financial or non-financial interests to disclose.

\newpage

\hypertarget{bibliography}{}

\end{document}